\documentclass[aps,pra,twocolumn,letterpaper,showpacs,superscriptaddress,floatfix,10pt]{revtex4-1}
\usepackage[pdftex]{graphicx}% Include figure files
\graphicspath{ {./images/} }
\usepackage{dcolumn}% Align table columns on decimal point
\usepackage{bm}% bold math
\usepackage{bbm}% bold math
\usepackage[usenames, dvipsnames]{color}
\usepackage{comment}
\usepackage[colorlinks, linkcolor=blue]{hyperref}

\usepackage{layouts}

\usepackage[T1]{fontenc}

\usepackage{amsfonts}
\usepackage{amssymb}
\usepackage{amsmath}
\usepackage{mathtools}
\usepackage{tensor}

\usepackage{amsthm}
\theoremstyle{remark}

\usepackage{multirow}
\usepackage[scr=boondoxo, scrscaled=1.05]{mathalfa}

\usepackage{subfigure}
\usepackage{overpic}

\usepackage{array}
\newcolumntype{L}[1]{>{\raggedright\let\newline\\\arraybackslash\hspace{0pt}}m{#1}}
\newcolumntype{C}[1]{>{\centering\let\newline\\\arraybackslash\hspace{0pt}}m{#1}}
\newcolumntype{R}[1]{>{\raggedleft\let\newline\\\arraybackslash\hspace{0pt}}m{#1}}
\usepackage{tabularx}

\def\KeyWord#1{$\backslash$\IfColor{$\!\!$\textRed{#1}\textBlack}{#1}$\!\!$}

\usepackage{wasysym}

\newcommand{\complexes}{\mathbb{C}}

\renewcommand{\d}{\mathrm{d}}

\def\bra#1{\langle#1|}
\def\ket#1{|#1\rangle}

\def\qexp#1#2{\bra{#2}#1\ket{#2}}

\begin{document}

\title{Phenomenology of the Prethermal Many-Body Localized Regime}

\author{David M. Long}
\email{dmlong@bu.edu}
\affiliation{Department of Physics, Boston University, Boston, Massachusetts 02215, USA}

\author{Philip J. D. Crowley}
\affiliation{Department of Physics, Massachusetts Institute of Technology, Cambridge, Massachusetts 02139, USA}

\author{Vedika Khemani}
\affiliation{Department of Physics, Stanford University, Stanford, California 94305, USA}

\author{Anushya Chandran}
\affiliation{Department of Physics, Boston University, Boston, Massachusetts 02215, USA}

\date{\today}

\begin{abstract}
The dynamical phase diagram of interacting disordered systems has seen substantial revision over the past few years.
Theory must now account for a large \emph{prethermal many-body localized} (MBL) regime in which thermalization is extremely slow, but not completely arrested.
We derive a quantitative description of these dynamics in short-ranged one-dimensional systems using a model of successive many-body resonances.
The model explains the decay timescale of mean autocorrelators, the functional form of the decay---a stretched exponential---and relates the value of the stretch exponent to the broad distribution of resonance timescales.
The Jacobi method of matrix diagonalization provides numerical access to this distribution, as well as a conceptual framework for our analysis.
The resonance model correctly predicts the stretch exponents for several models in the literature.
Successive resonances may also underlie slow thermalization in strongly disordered systems in higher dimensions, or with long-range interactions.
\end{abstract}

\maketitle

Localized systems fail to thermalize under their own dynamics due to strong spatial inhomogeneities~\cite{Anderson1958}. Without interactions, a stable fully localized phase exists in any dimension~\cite{Thouless1974,Abrahams2010,Segev2013a}. Including interactions, the existence of \emph{many-body localization} (MBL) becomes difficult to confirm. The consensus from the last decade and a half~\cite{Gornyi2005,Basko2006,Oganesyan2007,Pal2010,Imbrie2016b,deRoeck2017,Tikhonov2021} is that sufficiently strong random disorder produces stable MBL in short-ranged one dimensional systems only. Specifically, the best-studied model---the Heisenberg chain with random fields~\cite{Pal2010}---was believed to have a direct transition from a thermalizing phase to a fully MBL phase at a critical disorder strength \(W=W_c\) between 3 and 6 (in units of the Heisenberg coupling)~\cite{Pal2010,Luitz2015}.

\begin{figure}[h!]
    \centering
    \includegraphics[width=\linewidth]{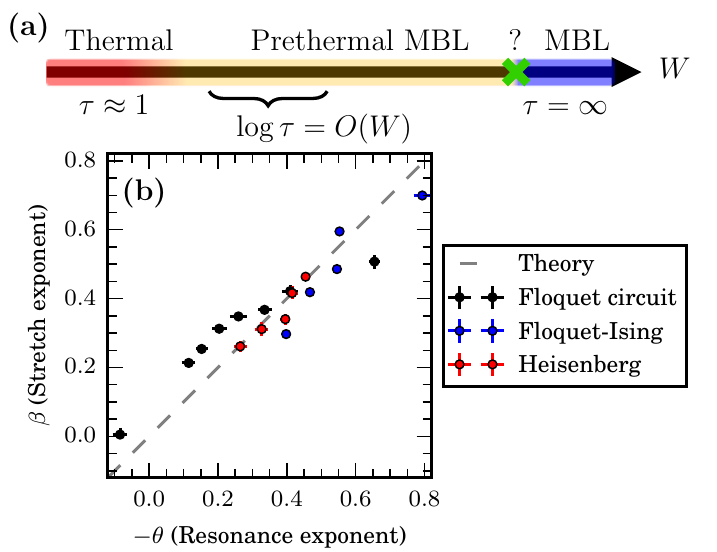}
    \caption{(\textbf{a}) Disordered many-body systems cross over from a well-thermalizing regime into a  \emph{prethermal many-body localized} regime, where the local equilibration time \(\tau\) grows exponentially with the disorder strength \(W\). Any transition to an MBL phase must occur at much larger disorder strength. (\textbf{b}) The successive resonance model predicts that the stretch exponent (\(\beta\)) appearing in stretched exponential decay of autocorrelators equals another exponent (\(-\theta\)) which describes the broad distribution of resonance timescales. Data from a one-dimensional Floquet circuit model of MBL~\cite{Morningstar2022}, a Floquet-Ising model~\cite{Lezama2019}, and the usual disordered Heisenberg model~\cite{Schiulaz2019,Lezama2021} are broadly consistent the prediction \(\beta = -\theta\) in the prethermal MBL regime.}
    \label{fig:phase_diagram}
\end{figure}

However, numerical evidence has been accumulating that this understanding is wrong---there is no transition near \(W = 3\)~\cite{Abanin2021,Suntajs2020,Schulz2020,Taylor2021,Sels2021,Sels2021b,Morningstar2022,Sels2021a,Sierant2022}. In fact, recent studies suggest \(W_c \gtrsim 20\)~\cite{Morningstar2022,Sels2021a}, which is larger than numerically or experimentally observable~\cite{Crowley2020,Morningstar2022}. The phase diagram must thus be modified to contain a large \emph{prethermal MBL} regime, in which the system appears to be localized for a long time~(\autoref{fig:phase_diagram}(a)).

Prethermal MBL phenomenology has been studied in short chains using exact diagonalization techniques~\cite{Suntajs2020,Sierant2020,Sels2021,Vidmar2021,Sels2021b}. Three key features have emerged: exponential growth of the thermalization time \(\tau\) with disorder, approximately logarithmic decay of autocorrelators up to a time \(O(\tau)\), and apparent localization when \(\tau\) exceeds the Heisenberg time \(\tau_{\mathrm{Heis}} = O(2^L)\) in finite chains. Rare regions of anomalously high disorder can neither explain the slow decay, nor are they empirically observed in this parameter regime~\cite{Schulz2020,Taylor2021} (Appendix~\ref{app:rareregion}). Rather, the observed decay~\cite{Gopalakrishnan2015,Agarwal2015,Agarwal2015b,BarLev2015,Luitz2016,Znidaric2016,Serbyn2017} and apparent localization can be partially explained~\cite{Crowley2020,Garratt2021} through \emph{resonances} between many-body states~\cite{Gopalakrishnan2015,Khemani2017,Villalonga2020,Crowley2020,Garratt2021}.

\begin{figure}
    \centering
    \includegraphics{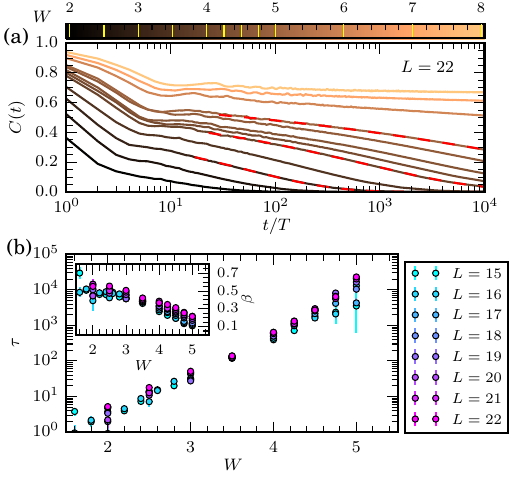}
    \caption{(\textbf{a}) Within a broad regime of disorder strengths (specific values marked yellow in the color bar), local autocorrelation functions~\eqref{eqn:Ct} for the Floquet circuit model decay very slowly. Fits to a stretched exponential (red, dashed for \(W \in \{3,4,5\}\)) show excellent agreement with the numerical data. (\textbf{b}) The decay times \(\tau\) extracted from stretched exponential fits~\eqref{eqn:stretched_exp} grow exponentially with disorder \(W\). (Fits with \(\tau \gtrsim 10^4\) exceed our maximum simulation time, and are unreliable.) (\textbf{Inset}) The stretch exponent \(\beta\) decreases with disorder, and increases weakly with system size (Appendix~\ref{subapp:fits}).}
    \label{fig:correlators}
\end{figure}

At short times, a product state may \emph{resonate} with another state with a locally different magnetization pattern. Physically, this manifests as large oscillations in local autocorrelators. In this Letter we propose that, in the prethermal regime, states become involved in still more resonances at longer times. Thus, a hierarchy emerges of \emph{successive resonances} forming at progressively longer timescales. Our statistical treatment of this resonance formation predicts exponentially long thermalization times, and stretched exponential decay of disorder averaged autocorrelators. Numerically, we find that autocorrelators indeed decay as a stretched exponential~\cite{Lezama2019}~(\autoref{fig:correlators}).

The Jacobi algorithm for iterative matrix diagonalization~\cite{Jacobi1846,Golub2000} is the basis of our analytical framework, and allows us to extract the distribution of resonance frequencies. The distribution is described by a power law with exponent \(-1+\theta\)~\cite{Crowley2020}. The successive resonance model predicts that the stretch exponent \(\beta\) for autocorrelator decay is linearly related to \(\theta\):
\begin{equation}
    \beta = - \theta.
    \label{eqn:beta_theta}
\end{equation}
Both our own numerics and previously published data show good agreement with this prediction (\autoref{fig:phase_diagram}(b)).

\emph{Dynamics of autocorrelators}.---Infinite-temperature autocorrelation functions are a measurable probe of thermalization, and their slow decay is a notable characteristic of the prethermal regime~\cite{Schreiber2015,Luitz2016b,Lezama2019,Lezama2021}. In a disordered model, \autoref{fig:correlators} shows that autocorrelators decay as a stretched exponential~\eqref{eqn:stretched_exp} with a decay constant which is exponential in the disorder strength~\eqref{eqn:logtau}.

We consider autocorrelators of operators which are diagonal in the disorder basis (the \(z\) basis). The numerics presented in the main text use
\begin{equation}
    C(t) = \frac{1}{2^L}[\mathrm{Tr}(\sigma_0^z(t) \sigma_0^z(0))],
    \label{eqn:Ct}
\end{equation}
in the Floquet circuit model of Ref.~\cite{Morningstar2022} with periodic boundary conditions, described in detail in Appendix~\ref{subapp:models}. Here, \(\sigma^z_0\) is the \(z\) spin operator on an arbitrary site (labelled \(0\)), \(L\) is the number of qubit degrees of freedom, \(\sigma^z_0(t) = U(t)^\dagger \sigma^z_0 U(t)\), \(U(t)\) is the unitary evolution operator, and square brackets denote a disorder average. In this model, every \(\sigma^z\) operator is conserved in the \(W \to \infty\) limit (Appendix~\ref{subapp:models}).

The Floquet circuit model has a well-thermalizing regime for \(W \ll 1\), where \(C(t)\) rapidly decays to zero. In any MBL phase, \(C(t)\) acquires a non-zero late time value. In the intermediate regime of prethermal MBL, \(1 \lesssim W \lesssim 25\), \(C(t)\) decays slowly to zero in the \(L \to \infty\) limit~\cite{Morningstar2022}.

In more detail, in the intermediate regime, \(C(t)\) first drops to some \(O(1)\) value within a few tens of periods, and then decays very slowly. The functional form of this decay appears logarithmic at small system sizes or short times~\cite{Sels2021}, but a better fit for larger system sizes is to a stretched exponential~(\autoref{fig:correlators}, Appendix~\ref{subapp:fits}),
\begin{equation}
    C(t) \sim A e^{-(t/\tau)^\beta}.
    \label{eqn:stretched_exp}
\end{equation}
(It is notoriously difficult to distinguish stretched exponential relaxation from a logarithm at intermediate times~\cite{Amir2011}.)

The timescale for decay (\(\tau\)) is extracted from a fit of this functional form to the late-time data for \(C(t)\). Consistent with other recent observations~\cite{Suntajs2020,Sierant2020,Sels2021}, \(\tau\) increases exponentially in the disorder strength.
\begin{equation}
    \log \tau = O(W).
    \label{eqn:logtau}
\end{equation}
In a Hamiltonian system, local equilibration on the timescale \(\tau\) would be followed by slow hydrodynamic decay.

The observations \eqref{eqn:stretched_exp} and \eqref{eqn:logtau} are the primary features that the model of successive resonances explains.

\emph{Jacobi algorithm}.---In the prethermal MBL regime, eigenstates of large systems should be expected to obey the eigenstate thermalization hypothesis (ETH)~\cite{Deutsch1991,Srednicki1994,DAlessio2016}. This makes them a poor basis for predicting finite time dynamics of local correlators. It is more revealing to use a short time expansion in a dressed basis.

The Jacobi algorithm for matrix diagonalization~\cite{Jacobi1846,Golub2000} provides a convenient numerical tool for constructing such a dressed basis. It also provides a more concrete framework within which to understand what is meant by a many-body resonance in a system which, ultimately, thermalizes.

We describe the algorithm for the case of a Hermitian operator (the Hamiltonian, \(H\)). Generalizations to the unitary case~\cite{Goldstine1959,Ruhe1967} are appropriate for the Floquet setting, and are discussed in Appendix~\ref{subapp:jacobi}.

The algorithm begins by identifying the largest (in absolute value) off-diagonal matrix element of \(H\), \(H_{ab}\), in the \(z\)-basis. The \(2\times 2\) block containing this element is diagonalized by the unitary rotation \(R_0\). (Note that \(R_0\) affects the entire \(a\) and \(b\) rows and columns of \(H\).) The \((a,b)\) element of \(H\) is then set to zero in the rotated matrix \(H(\Gamma_1) = R_0^\dagger H R_0\), where \(\Gamma_1\) is a flow time for the algorithm, defined below in Eq.~\eqref{eqn:flowtime}. We say the element \(H_{ab}\) of \(H\) is decimated, in analogy to the renormalization group.

This process is iterated, so that the weight in the off-diagonal of \(H(\Gamma_{j+1}) = R_{j}^\dagger H(\Gamma_{j}) R_{j}\) strictly decreases. This procedure constructs a basis
\begin{equation}
    \ket{a(\Gamma_j)} = R_{j-1} \cdots R_1 R_0 \ket{a(\Gamma_0)}
\end{equation}
(where \(\{\ket{a(\Gamma_0)}\}\) is the bare product state basis) which is dressed by the fast degrees of freedom in \(H\).

The flow time \(\Gamma\) is defined in terms of a physical timescale associated to the basis \(\{\ket{a(\Gamma)}\}\) (\(\hbar=1\)),
\begin{equation}
    \frac{2\pi}{\Gamma} = \sqrt{\frac{1}{L 2^L}\sum_{a \neq b} |\bra{b(\Gamma)}H\ket{a(\Gamma)}|^2}.
    \label{eqn:flowtime}
\end{equation}
and strictly increases throughout the course of the algorithm~\cite{Golub2000}. Henceforth, we neglect all sub-exponential factors of \(L\), as in the denominator of Eq.~\eqref{eqn:flowtime}.

The Jacobi algorithm diagonalizes \(H\) within \(O(4^L)\) steps. Only \(O(W 2^L)\) steps are necessary to construct the \emph{dressed basis} useful for computing autocorrelators. The dressed states only have large overlap with \(O(1)\) bare states on average, as discussed in the next section.

\emph{Dynamics of successive resonance}.---Expressing the autocorrelator \(C(t)\) in the dressed basis relates it to the statistics of the Jacobi algorithm~\eqref{eqn:Ct_char_fn}. With two natural assumptions---that dynamics are dominated by \emph{sparse resonances}~\eqref{eqn:Jacobi_res_step}, and that the timescales associated to these resonances are power law distributed~\eqref{eqn:rho_plaw}---stretched exponential decay follows.

When calculating autocorrelators of some operator \(Z\) (assumed to be diagonal in the \(\{\ket{a(\Gamma_0)}\}\) basis) for \(t \ll \Gamma\), we can treat the Hamiltonian as being diagonal in the basis \(\{\ket{a(\Gamma)}\}\) at the cost of introducing a well-controlled error:
\begin{align}
    C_Z(t) &= \frac{1}{2^L}[\mathrm{Tr}(Z(t) Z(0))] \nonumber\\
    &= \frac{1}{2^L} \left[\sum_{a, b} |Z_{ab}(\Gamma)|^2 e^{-i \omega_{ab}(\Gamma) t}\right] + O((t/\Gamma)^2),
    \label{eqn:correlator_approx}
\end{align}
where \(Z_{ab}(\Gamma) = \bra{a(\Gamma)} Z \ket{b(\Gamma)}\), \(\omega_{ab}(\Gamma) = \bra{b(\Gamma)} H \ket{b(\Gamma)} - \bra{a(\Gamma)} H \ket{a(\Gamma)}\), and square brackets are again used to denote a disorder average.

The joint distribution function of \(|Z_{ab}(\Gamma)|^2\) and \(\omega_{ab}(\Gamma)\), \(p(Z^2,\omega;\Gamma)\), determines \(C_Z(t)\) through
\begin{equation}
    C_Z(t) = 2^L [Z^2 e^{-i\omega t}]_{p(\Gamma)} +O((t/\Gamma)^2).
    \label{eqn:Ct_dist}
\end{equation}
The subscript on the square brackets indicates the distribution over which the average is performed.

We can deduce properties of \(p\), and hence \(C_Z(t)\), from the Jacobi algorithm. Namely, that large matrix elements in the distribution only arise due to occasional large rotations in the Jacobi algorithm.

As \(Z\) is diagonal in the initial basis \(\{\ket{a(\Gamma_0)}\}\), its off-diagonal elements only become large when some rotation \(R_k\) affecting that element is also large. This happens when the decimated off-diagonal matrix element is much larger than the difference in diagonal elements \(\omega_{ab}(\Gamma_{k})\)---that is, when the states \(\ket{a(\Gamma_{k})}\) and \(\ket{b(\Gamma_{k})}\) are \emph{resonant}. Then, in the next round of iteration,
\begin{align}
    |Z_{ab}(\Gamma_{k+1})|^2 &=O(1), \nonumber\\
    H_{aa}(\Gamma_{k+1}) &\approx H_{aa}(\Gamma_{k}) \pm |H_{ab}(\Gamma_{k})|,
    \label{eqn:Jacobi_res_step}
\end{align}
and similarly \(H_{bb}(\Gamma_{k+1}) \approx H_{bb}(\Gamma_{k}) \mp |H_{ab}(\Gamma_{k})|\).

We make the approximation that rotations are either trivial or cause resonances~\eqref{eqn:Jacobi_res_step}~\cite{Crowley2020}. Only the resonances produce dynamics.

\begin{figure}
    \centering
    \includegraphics{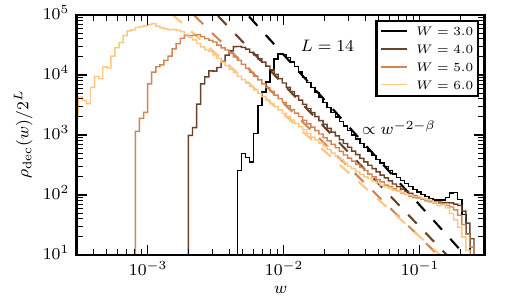}
    \caption{The Jacobi decimated elements for the Floquet circuit model (the main text discusses the Hamiltonian case) for \(200 \cdot 2^L\) iterations are approximately power law distributed for intermediate decimated weights \(w_{ab}\), which generalize the decimated matrix elements \(|H_{ab}|\) to the unitary case (Appendix~\ref{subapp:jacobi}). Furthermore, the power law is in good agreement with the predicted \(-2-\beta\) from the successive resonance model (dashed lines). For this number of iterations, the average number of resonances per state is \(\lesssim 1\). Matrix element distributions are averaged over 100 disorder realizations, \(\rho_{\mathrm{dec}}\) is normalized as a number density, and \(\beta\) is fit from \autoref{fig:correlators}.}
    \label{fig:mat_dist}
\end{figure}

Before thermalization, resonances are \emph{sparse}. The probability of a resonance occurring in a given rotation is small, \(P(|\omega_{ab}| < |H_{ab}|) = O(W^{-1})\). Further, the prefactor hidden in this scaling expression is also small: between \(0.1\%\) and \(1\%\) of rotations are resonances in the studied parameter regimes. Thus, after \(O(W)\) Jacobi steps per state (as in \autoref{fig:mat_dist}), every dressed state \(\ket{a(\Gamma)}\) is involved in \(O(1)\) resonances on average.

Technically, the sparse resonance assumption is that \(|Z_{ab}|^2 = O(1)\) only for resonant states. This ignores the effects of successive resonances which may reduce \(|Z_{ab}|^2\), and the possibility of many small rotations producing a large \(|Z_{ab}|^2\). This assumption is valid provided that the number of resonances per state is \(O(1)\). This provides a large intermediate window, a few multiples of \(\tau\), in which we can make predictions.

The resonance assumption splits \(p\) into a part due to resonances, which contributes to \(C_Z(t)\), and a part where matrix elements are all close to zero:
\begin{equation}
    p(Z^2,\omega;\Gamma) \approx \delta(Z^2) p_0(\omega;\Gamma) + p_{\mathrm{res}}(Z^2, \omega;\Gamma).
\end{equation}

Equation~\eqref{eqn:Jacobi_res_step} leads to two conclusions regarding \(p_{\mathrm{res}}\). First, the matrix element \(|Z_{ab}(\Gamma_k)|^2\), being \(O(1)\), does not depend strongly on \(\omega_{ab}(\Gamma_{k})\). Consequently, the expectation of \(Z^2\) at fixed \(\omega\) in \(p_{\mathrm{res}}\) can be factorized out of Eq.~\eqref{eqn:Ct_dist}. This  gives a key intermediate result:
\begin{equation}
    C_Z(t) 
    \approx 2^L[Z^2]_{p_{\mathrm{res}}(\Gamma)} \mathcal{F}\{p_{\mathrm{res}}(\omega;\Gamma)\}(t),
    \label{eqn:Ct_char_fn}
\end{equation}
where \(p_{\mathrm{res}}(\omega;\Gamma)\) is the marginal distribution function of the resonance frequencies and \(\mathcal{F}\{\cdot\}\) is the Fourier transform.

The second consequence is found by repeatedly applying Eq.~\eqref{eqn:Jacobi_res_step} to find the energy differences \(\omega\). They are of the form 
\begin{equation}
    \omega_{ab}(\Gamma) = \sum_{\Gamma_k<\Gamma} \mu_k |H(\Gamma_k)|,
    \label{eqn:omega}
\end{equation}
where \(\mu_k = \pm 1\), and the sum runs over matrix elements \(|H(\Gamma)|\) responsible for a resonance in either state \(\ket{a(\Gamma)}\) or \(\ket{b(\Gamma)}\) at flow time \(\Gamma\). We have neglected the initial value \(\omega_{ab}(\Gamma_0)\), which must be small if the states are to become resonant. Eq.~\eqref{eqn:omega} encodes the effect of many resonances, each contributing to dynamics at progressively longer timescales \(2\pi/|H(\Gamma_k)|\).

Equation~\eqref{eqn:Jacobi_res_step} relates the frequencies \(\omega_{ab}(\Gamma)\) to the resonance timescales, and hence the distribution of decimated elements. Our central assumption, verified numerically in~\autoref{fig:mat_dist}, is that the distribution of decimated elements is a \emph{power law} (c.f. Ref.~\cite{Vidmar2021}). This is natural if the dressed basis is quasilocal, as the distribution of matrix elements of a quasilocal operator in a quasilocal basis is a power law in one dimension~\cite{Crowley2020,Morningstar2022,Garratt2021,Garratt2022}. (Matrix elements decrease exponentially with spatial range, but there are exponentially many of them.)

As \(\omega_{ab}(\Gamma)\) is the sum of many independent variables, the central limit theorem may be invoked. The limit distribution for a sum of power law distributed variables is not normal, but is rather a \emph{L\'evy stable distribution}~\cite{Uchaikin2011}. The Fourier transform of a L\'evy distribution is a stretched exponential, which leads to the observed form of the decay~\eqref{eqn:stretched_exp} through Eq.~\eqref{eqn:Ct_char_fn}.

The distribution of decimated elements is parameterized as a power law ansatz with an exponent \(\theta\)~\cite{Crowley2020}:
\begin{equation}
    \rho_{\mathrm{dec}}(|H|) = 2^L C |H|^{-2+\theta}.
    \label{eqn:rho_plaw}
\end{equation}
Explicitly reinserting a local energy scale \(J = O(\Gamma_0^{-1})\), dimensional analysis gives \(C=O(J^{1-\theta})\). The distribution of \(|H(\Gamma_k)|\) involved in resonances (treating \(\omega_{ab}(\Gamma)\) and \(|H_{ab}(\Gamma)|\) as uncorrelated) is
\begin{multline}
    \rho_{\mathrm{res}}(|H|) = P(|\omega| < |H|) \rho_{\mathrm{dec}}(|H|) \\
    \approx 2^{L+1} p(\omega = 0) C |H|^{-1+\theta},
    \label{eqn:rho_res}
\end{multline}
where we assumed \(|H|\) is small, so that \(P(|\omega| < |H|) \approx  2 p(\omega = 0) |H|\), and \(p(\omega) = O(W^{-1})\) is the \(\omega\) marginal of \(p(Z^2,\omega)\).

The distribution of resonances~\eqref{eqn:rho_res} is also a power law, but with a larger exponent, \(-1 + \theta\), than \(\rho_{\mathrm{dec}}\).

The exponent \(\theta\) appearing in Eqs.~\eqref{eqn:rho_plaw} and~\eqref{eqn:rho_res} is the central parameter of the \emph{single} resonance model introduced in Ref.~\cite{Crowley2020} (see also Ref.~\cite{Gopalakrishnan2015}). With the chosen parameterization, \(\theta <0\) implies thermalization. Successive resonances may cause a drift of \(\theta\) with \(\Gamma\). However, for sufficiently negative \(\theta\), the system thermalizes before any significant drift, and \(\theta\) may be treated as a constant. \autoref{fig:mat_dist} shows this is a reasonable approximation in accessible parameter regimes.

The exponent \(\theta\) also controls the distribution of matrix elements of generic local operators in the \(\{\ket{a(\Gamma)}\}\) basis, not just the decimated elements~\cite{Garratt2022} (Appendix~\ref{app:offdiag}).

Appendix~\ref{app:multires} computes the Fourier transform~\eqref{eqn:Ct_char_fn} and shows that, for \(\theta < 0\),
\begin{equation}
    C_Z(t) \sim A e^{-(t/\tau)^{-\theta}}
    \quad
    \text{for}
    \quad
    J^{-1} \ll t \ll \omega_c^{-1},
    \label{eqn:Ct_prediction}
\end{equation}
where \(A\) is a constant, \(\omega_c = O(\tau^{-1})\) is a small frequency cutoff, and \(J\) gives the large frequency cutoff.

The linear scaling of \(\log(J\tau)\) with \(W/J\) follows from our previous assumptions and a linearization of \(-\theta^{-1}\) in \(W/J\)~\cite{Crowley2020} (Appendix~\ref{app:multires}). The power law form of \(\rho_{\mathrm{res}}\) is appropriate while each state is involved in few resonances. It breaks down when
\begin{equation}
    \int_{\omega_c}^J \rho_{\mathrm{res}}(H)\, \d H = O(2^L),
\end{equation}
which immediately provides
\begin{equation}
    \log(J \tau) = O(-\theta^{-1}\log(-\theta W/J))
\end{equation}
(using \(p(\omega=0) = O(W^{-1})\), \(C J^{\theta} = O(J)\), and that 
the ratio \(J/\omega_c\) is proportional to \(J\tau\)). The exact dependence of \(-\theta^{-1}\) on \(W/J\) is unknown. Nevertheless, as \(-\theta^{-1}\) is slowly varying in the numerically accessible \(W/J\) range (inset of \autoref{fig:correlators}(b)), we linearize it. This gives \(\log(J\tau) = O(W/J)\).

The dependence of \(\log(J \tau)\) may be different for larger \(W/J\). The purely linear form for \(\log(J \tau)\) obtained numerically (\autoref{fig:correlators}(b)) thus remain unexpected in the successive resonance model.

Equation~\eqref{eqn:Ct_prediction} accounts for both of our main observations in \autoref{fig:correlators}. Furthermore, we have arrived at a falsifiable prediction: the stretch exponent \(\beta\) of the decay should be given by \(\beta = - \theta\)~\eqref{eqn:beta_theta}. An independent numerical measurement of \(\theta\) can test this prediction. Indeed, fitting a power law \(-2 + \theta\) to the distribution of decimated elements in \autoref{fig:mat_dist} produces exponents \(-\theta\) which are in broad agreement with \(\beta\) as fit to \(C(t)\)~(\autoref{fig:phase_diagram}(b)). This agreement extends to several models with prethermal MBL regimes~\cite{Lezama2019,Lezama2021,Schiulaz2019}. Note that, as \(\rho_{\mathrm{dec}}\) is not a pure power law, agreement should not be exact.

\emph{Discussion}.---Theory forbids a stable MBL phase in many settings in which experiment and numerics observe MBL phenomenology~\cite{Wahl2019,Rubio2019,Chertkov2021,vanHorssen2015,Yao2016,deRoeck2016,LeBlond2021,Bulchandani2022,Burin2006,Yao2014,Smith2016,Nandy2022}. Even in one dimension, MBL may only exist at much larger disorder strengths than previously anticipated~\cite{deRoeck2017,Crowley2022b,Morningstar2022,Sels2021,Sels2021b}.
These settings are instances of prethermal MBL. We have begun the study of such prethermal dynamics in one dimension. The successive resonance model predicts stretched exponential decay of autocorrelators, an exponentially long thermalization time, and the value of the stretch exponent, all of which are supported by numerics in several models at intermediate values of \(W\). These values of \(W\) lie in the regime that diagonalization studies previously identified as critical.  

At larger values of \(W\) in the prethermal regime (previously identified as many-body localized), the decimated elements can be controlled by a power law exponent \(\theta\) that is positive for small Jacobi flow times. Here, the flow of \(\theta\) cannot be ignored, as thermalization should be signaled by a negative \(\theta\). Testing resonance model predictions then requires longer flow times and larger system sizes, or an analytic theory of the flow of \(\theta\).

The successive resonance model does not use any properties of the random potential. As such, its predictions are identical for correlated potentials in one dimension. (To compare disorder strengths between random and correlated potentials, the distribution of energy differences between neighboring sites should be the same, not the distribution of energies on a site~\cite{Khemani2017}.)

More generally, successive resonance model predictions are identical in any setting where \(\rho_{\mathrm{dec}}\) is a power law with \(\theta < 0\). Future work should check this in higher dimensions~\cite{Wahl2019,Rubio2019,Chertkov2021}, translationally-invariant MBL~\cite{vanHorssen2015,Yao2016}, and with long range interactions~\cite{Burin2006,Yao2014,Smith2016,Nandy2022}. Whenever \(\rho_{\mathrm{dec}}\) has a strong separation of scales, the model may still be predictive, although decay need not be stretched exponential. An interesting open question is the applicability to Floquet prethermalization~\cite{Mori2018}, in which there is only one long-lived global conservation law, but fidelities still show slow decay~\cite{ODea2023}.

In Anderson models on random regular graphs and related random matrix models, return probabilities exhibit stretched exponential decay~\cite{Monthus2017,Bera2018,DeTomasi2020,Khaymovich2020,Khaymovich2021,Colmenarez2022,Tikhonov2021}. With the assumption of sparse resonances, the formal calculations in these models are very similar to ours. The application of the Jacobi method to random regular graphs is worth exploring.

The Jacobi algorithm provides an effective off-diagonal matrix element distribution at different time scales. Its applications to quantum dynamics, the emergence of hydrodynamics, and connections to other techniques~\cite{Pekker2017,Bravyi2011,delaLlave2001,Agarwal2015b} deserve further investigation. Indeed, the most rigorous analysis of MBL uses the Jacobi algorithm~\cite{Imbrie2016a,Imbrie2016b}.

\emph{Acknowledgements}.---The authors are particularly grateful to L. F. Santos for providing high quality autocorrelator data for the Heisenberg model, and to A. Morningstar for code related to the Floquet circuit model. We also thank D. Huse, I. M. Khaymovic, Y. H. Kwan, C. Laumann, A. Morningstar, A. Polkovnikov, D. Sels, and T. Veness for helpful discussions and related collaborations. This work was supported by: NSF Grant No. DMR-1752759, and AFOSR Grant No. FA9550-20-1-0235 (D.L. and A.C.); the NSF STC ``Center for Integrated Quantum Materials'' under Cooperative Agreement No. DMR-1231319 (P.C.); and by NSF Grant No. NSF PHY-1748958 (KITP).
V.K. acknowledges support from the US Department of Energy under Early Career Award No. DE-SC002111, a Sloan Research Fellowship and Packard Fellowship. Numerical work was performed on the BU Shared Computing Cluster.

\bibliography{prethermalMBL}

\clearpage

\appendix
\section{Irrelevance of Rare-Regions Effects at Intermediate Disorder}
    \label{app:rareregion}

    This appendix summarizes results from the recent literature which indicate that rare-region effects are not the dominant mechanism determining dynamics---including decay rates and functional forms of autocorrelators---in the regime of disorder strengths and timescales we study in this work.

    The many-body localization transition is believed to be controlled by \emph{rare-region effects}~\cite{Gopalakrishnan2015,Agarwal2015,deRoeck2017}. With random disorder, there are generically sparse patches of the system where the disorder configuration happens to be much more (or less) localizing than typical for a given disorder strength. In the thermal phase these rare regions serve as bottlenecks slowing the spread of correlations, while in the localized phase they act as local ``baths'' for the surrounding degrees of freedom. In both phases, they are believed to play a crucial role in determining long-time dynamics at disorder strengths close to the critical value. We refer the interested reader to the review in Ref.~\cite{Agarwal2017}.

    This work concerns disorder strengths historically identified as being a part of the critical fan near the MBL transition~\cite{Pal2010,Luitz2015}. As such, dynamics in the regime of disorder and system sizes we study has previously been explained by appealing to the effects of rare regions~\cite{Gopalakrishnan2015,Agarwal2015,Gopalakrishnan2016,Agarwal2017}. However, recent results indicate that rare regions are not a significant factor determining dynamics at intermediate disorder strengths and timescales.

    While rare-region effects are still believed to play a crucial role in the asymptotic MBL transition, recent numerical results indicate that this transition occurs at a much higher value of disorder strength than previously estimated~\cite{Suntajs2020,Schulz2020,Taylor2021,Sels2021,Sels2021b,Morningstar2022,Sels2021a,Sierant2022}. Early numerical investigations fit critical exponents \(\nu\) far smaller than that predicted by rare-region based renormalization group schemes~\cite{Pal2010,Luitz2015,Vosk2015,Potter2015,Zhang2016,Goremykina2019,Dumitrescu2019} (indeed, too small to be related to an asymptotic critical point~\cite{Chandran2015}). Strong drifts in the estimated critical disorder with system size also indicated that the asymptotic critical point was misplaced~\cite{Pal2010,Luitz2015,Suntajs2020}. Lower bounds of the critical disorder strength in the random-field Heisenberg model are now placed near \(W_c \gtrsim 20\)~\cite{Morningstar2022,Sels2021a}.

    Thus, there is not a clear analytical motivation to suspect rare-region physics near \(W \approx 3\), where dynamics is slow, but which is likely well outside any critical fan from an asymptotic transition. Of course, the possibility remains that such a critical fan happens to be extremely large, and \(W\approx 3\) still shows signs of rare-region effects. To address this speculation, one must turn to numerical evidence.

    There is an accumulating body of numerical evidence which indicates that rare-region effects are not relevant at intermediate disorder. Notably, Ref.~\cite{Schulz2020} studied the predictions of rare regions in disordered chains by computing the distribution of their resistivities. Those authors found that the broad distributions of resistivity predicted by rare-region effects were absent. This is the most direct numerical falsification of rare-region predictions at intermediate disorder known to the authors.

    There is a much larger body of indirect evidence against rare-region effects driving thermalization at intermediate disorder. This includes the numerical evidence showing that the asymptotic MBL transition is at a larger disorder strength than estimated at small system sizes~\cite{Suntajs2020,Schulz2020,Taylor2021,Sels2021,Sels2021b,Morningstar2022,Sels2021a,Sierant2022}.

    There is no obvious mechanism for rare-region effects to emerge in chains with quasiperiodic disorder, and the nature of the MBL transition in quasiperiodically disordered chains is predicted to differ from the randomly disordered case~\cite{Khemani2017b,Agrawal2020}. However, numerics at available system sizes at intermediate disorder show similar behavior in both these cases~\cite{Khemani2017b}. For instance, eigenstate-to-eigenstate variation in half-cut entanglement entropies are comparable to sample-to-sample variations. This indicates that the dominant mechanism of dynamics at such \(W\) must be present both in random systems and quasiperiodic systems.

    \begin{figure}
        \centering
        \includegraphics[width=\linewidth]{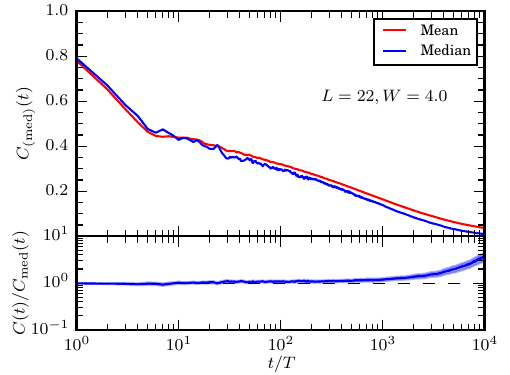}
        \caption{\label{fig:mean_med}The mean and median (over disorder) autocorrelators \(C(t)\) in the Floquet circuit model~\eqref{eqn:model} remain very close---within slightly more than one standard deviation (shaded region in lower panel)---for timescales less than and comparable to \(\tau \approx 10^3 T\). They begin to separate only well beyond \(\tau\).}
    \end{figure}

    A prediction of rare-region effects is that the mean (over disorder) autocorrelator \(C(t)\) should asymptotically separate from the median autocorrelator,
    \begin{equation}
        C_{\mathrm{med}}(t) = \frac{1}{2^L}[\mathrm{Tr}(Z(t)Z(0))]_{\mathrm{med}},
    \end{equation}
    where \([\cdot]_{\mathrm{med}}\) denotes a median over the disorder distribution. The mean \(C(t)\) is predicted to decay as a power law, while the median is predicted to decay as a stretched exponential~\cite{Gopalakrishnan2016}. Our data for the mean is consistent with stretched exponential decay (Appendix~\ref{subapp:fits}), contrary to the predictions of rare-region effects. Indeed, we only see signs of separation between the mean and median after many decay times \(\tau\) (\autoref{fig:mean_med}), indicating that rare-region effects do not explain the decay of autocorrelators at timescales where local equilibrium is still being established. This is the process we advocate proceeds primarily through resonances.

    In fact, there is a growing body of numerical evidence and analytic arguments which attribute thermalization at intermediate disorder to many-body resonances~\cite{Gopalakrishnan2015,Khemani2017,Villalonga2020,Crowley2020,Garratt2021}. This provides an alternative mechanism for thermalization, without appealing to rare-region effects.

\section{Numerical Procedures}

    This appendix describes the details of the numerical computations of autocorrelators (Appendix~\ref{subapp:dynamics}), the partial diagonalization of the Floquet operator using the Jacobi algorithm (Appendix~\ref{subapp:jacobi}), and the stretched exponential fits to numerically evaluated autocorrelators (Appendix~\ref{subapp:fits}). The different models we analyze are described in Appendix~\ref{subapp:models}.
    
    \subsection{Models}
        \label{subapp:models}
        
        The predictions of this Letter are generic for any one-dimensional strongly inhomogeneous model. To verify this explicitly, we have analyzed several different models used in the MBL literature.
        
        The model of primary focus is the Floquet-circuit model of Ref.~\cite{Morningstar2022}. This model is designed to be a clean numerical test bed for MBL---it has no conservation laws which would complicate relaxation to the infinite temperature state, and the interactions responsible for thermalization have no preferred local basis in which relaxation is particularly fast or slow. On a technical level, it is also easy to simulate dynamics in this model for large system sizes.

        The model describes periodic unitary dynamics of a length \(L\) qubit chain. The Floquet operator is decomposed into an on-site part and a nearest-neighbour interaction, \( U_F = U_{\mathrm{int}} U_{0}\).
        The on-site part consists of a tensor product of Haar-random single-site unitaries, \(U_{0} = \prod_{j=0}^{L-1} d_j\).
        The nearest neighbour interaction consists of two-site gates \(u_{j,j+1}\) applied in a random order,
        \begin{equation}
            U_{\mathrm{int}} = \prod_{k=0}^{L-1} u_{\pi(k), \pi(k)+1},
            \quad
            u_{j,j+1} = \exp[i M_{j,j+1}/W].
            \label{eqn:model}
        \end{equation}
        Here, \(M_{j,j+1}\) is a random matrix drawn from the \(4 \times 4\) Gaussian unitary ensemble (GUE), and is normalized so that \([\mathrm{Tr}(M_{j,j+1}^2)] = 2\), where square brackets denote an ensemble average. The randomly selected permutation \(\pi \in S_L\) determines the order of application of the gates, and the system has periodic boundary conditions.
        
        Fixing a basis on each site which diagonalizes \(d_j\), the operator with matrix \(\sigma^z_j\) in this basis is conserved by \(U_F\) in the limit \(W \to \infty\). The autocorrelator \(C(t)\) defined in the main text uses \(\sigma^z_0\).
        
        This model was characterized extensively in Ref.~\cite{Morningstar2022}. Briefly, level spacing ratios of \(L=20\) systems show a crossover from GUE statistics to Poisson statistics at \(W \approx 6\). Meanwhile, a dynamical phase transition due to avalanches must occur above \(W_c \gtrsim 25\).
        
        The second model we studied is the disordered Heisenberg model,
        \begin{equation}
            H = J\left(\sum_{j=0}^{L-1} h_j S^z_j + \vec{S}_j \cdot \vec{S}_{j+1}\right),
            \quad h_j \in [-W, W]
            \label{eqn:Heis_model}
        \end{equation}
        (in the zero magnetization sector) which is the standard model for MBL used in the literature. Autocorrelator data for this model was kindly provided to us by the authors of Refs.~\cite{Schiulaz2019,Lezama2021} for disorder values in the range \(W \in [1.25,2.5]\). We have also performed our own simulations (\autoref{fig:Ct_Heis}). Spectral statistics in this model for \(L \approx 20\) crosses over from the random matrix value to Poisson near \(W \approx 3\).
        
        Lastly, we investigate another model of Floquet MBL---the Floquet-Ising model of Ref.~\cite{Zhang2016},
        \begin{align}
            U_F &= e^{-i H_x T} e^{-i H_z T} \nonumber \\
            H_x &= \sum_{j = 0}^{L-1} g \gamma \sigma^x_j \nonumber \\
            H_x &= \sum_{j = 0}^{L-1} \sigma^z_j \sigma^z_{j+1} + (h + g\sqrt{1-\gamma^2}G_j)\sigma^z_j,
            \label{eqn:FIsi_model}
        \end{align}
        where \((g, h, T) = (0.9045, 0.8090, 0.8)\) and each \(G_j\) is an independent standard normal random variable. Autocorrelator data for this model is retrieved from Ref.~\cite{Lezama2019} for parameter values \(\gamma \in [0.4,0.6]\).
    
    \subsection{Dynamics}
        \label{subapp:dynamics}
        
        This section describes the computation of the autocorrelator \(C(t)\) in the Floquet circuit model~\eqref{eqn:model} and the Heisenberg model~\eqref{eqn:Heis_model}. Finite time dynamics in these models can be simulated for system sizes of over \(L = 20\) with modest computational resources, as we explain below.
        
        Due to the locality of the quantum circuit defining the Floquet circuit model, its Floquet operator can be applied to a state in the \(2^L\) dimensional Hilbert space with a computational cost of \(O(L 2^L)\). This allows the computation of autocorrelators \(C(t)\) for system sizes as large as \(L=22\) and integration times of \(t = 10^4 T\). Larger system sizes are certainly feasible with further computational effort.
        
        The computation of the trace in \(C(t)\) uses the typicality of expectation values in Haar random states~\cite{Lezama2019,Steinigeweg2014,Weisse2006}. Selecting a state \(\ket{\psi}\) uniformly at random in the Hilbert space, the correlator in this state
        \begin{equation}
            C_{\psi}(n T) = \bra{\psi} \sigma^z_0(nT)\sigma^z_0(0)\ket{\psi}
            = \bra{\psi} U_F^{\dagger n} \sigma^z_0 U_F^{n}\sigma^z_0\ket{\psi}
            \label{eqn:Cpsi}
        \end{equation}
        can be found as the matrix element of \(\sigma_0^z\) between \(U_F^{n}\ket{\psi}\) and \(U_F^{n}\sigma^z_0\ket{\psi}\). The expectation value~\eqref{eqn:Cpsi} in a random state is close to the trace, with a small variance,
        \begin{equation}
            C_\psi(t) = \frac{1}{2^L}\mathrm{Tr}(\sigma^z_0(t)\sigma^z_0(0)) + O(2^{-L/2}).
        \end{equation}
        
        Simulating dynamics in a single Haar random state provides an unbiased and low variance estimator for the infinite temperature autocorrelator. Averaging \(C_{\psi}(t)\) over random disoder realizations, with independently selected Haar random initial states, gives a similarly unbiased estimate of \(C(t)\).

\begin{figure}
            \includegraphics[width=\linewidth]{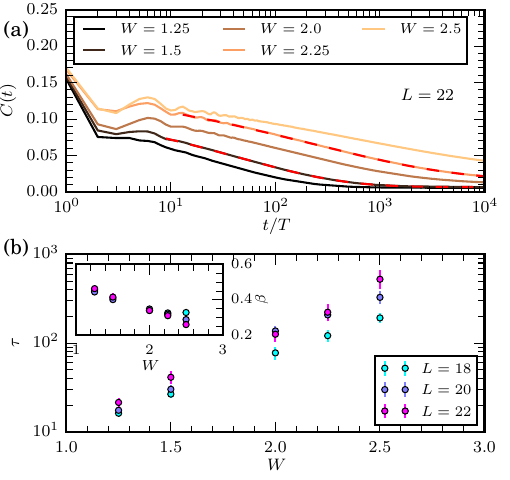}
            \caption{\label{fig:Ct_Heis}(\textbf{a}) As in the Floquet circuit model of the main text, local autocorrelation functions for the Heisenberg model decay very slowly. Fits to a stretched exponential plus a constant (red, dashed) show excellent agreement with the numerical data. (\textbf{b}) The decay times \(\tau\) extracted from the fits grow exponentially with disorder \(W\). (\textbf{Inset}) The stretch exponent \(\beta\) decreases with disorder.}
        \end{figure}

        Similarly, in the case of the Heisenberg model, \(U(t)\ket{\psi}\) and \(U(t) S^z_0 \ket{\psi}\) can be efficiently computed using Krylov space methods, providing an efficient computation of
        \begin{equation}
            C_{\psi}(t) = \bra{\psi} S^z_0(t)S^z_0(0)\ket{\psi}
            = \bra{\psi} U^{\dagger}(t) S^z_0 U(t) S^z_0\ket{\psi}
            \label{eqn:Cpsi}
        \end{equation}
        in the zero magnetization sector. The numerically calculated \(C(t)\) for the Heisenberg model, found by averaging \(C_{\psi}(t)\) over Haar random states \(\ket{\psi}\) and disorder configurations, is shown in \autoref{fig:Ct_Heis}. Due to the strong finite size effects in this energy conserving model, there is a significant satrutation value in the late-time \(C(t)\). We include this value as an additional phenomenological parameter when fitting \(C(t)\) to a stretched exponential (Appendix~\ref{subapp:fits}).

    \subsection{Jacobi Rotations}
        \label{subapp:jacobi}
        
        The Jacobi algorithm was first proposed in 1846~\cite{Jacobi1846}, and by this point has many generalizations and variations~\cite{Golub2000,Goldstine1959}. This section describes the specific instance of the Jacobi algorithm we use for Floquet operators (the Hermitian version is described in the main text), and the method of extracting the estimates of \(\theta\) shown in \autoref{fig:phase_diagram}(b). 
        
        The Jacobi algorithm for a general normal (equivalently, unitarily diagonalizable) matrix is similar in design to the Hermitian case described in the main text. The algorithm must contend with an additional complication due to the fact that a \(2 \times 2\) sub-block of a normal matrix is typically not itself diagonalizable. When the sub-block is not diagonalizable, the weight in the off-diagonal can only be minimized to a non-zero value by a rotation, rather than being completely eliminated.
        
        Given a two-dimensional complex matrix \(M = m_0 \mathbbm{1} + \vec{m} \cdot \vec{\sigma}\), where \(m_0 \in \complexes\) and \(\vec{m} \in \complexes^3\), we need the unitary \(R\) which minimizes the off-diagonal weight \(|M'_{01}|^2 + |M'_{10}|^2\) in
        \begin{equation}
            M' = R^\dagger M R.
        \end{equation}
        The solution is
        \begin{equation}
            R(\vec{m}) = 
            \begin{pmatrix}
                \cos(\eta/2) & -e^{-i \phi} \sin(\eta/2) \\
                e^{i \phi} \sin(\eta/2) & \cos(\eta/2)
            \end{pmatrix},
            \label{eqn:normal_R}
        \end{equation}
        where
        \begin{multline}
            (\cos \phi \sin \eta, \sin \phi \sin \eta, \cos\eta) 
            \propto \mathrm{Re}\left[ \vec{m}/\sqrt{\vec{m}\cdot\vec{m}} \right].
        \end{multline}
        Here, \(\vec{m}\cdot\vec{m}\) is the symmetric (\emph{not} conjugate symmetric) dot product and either branch of the complex square root may be chosen. Different choices of sign result in rotations \(R(\vec{m})\) related by permuting states.
        
        With these expressions in hand, the algorithm proceeds largely as described for the Hermitian case. Given a normal matrix \(M\), one identifies the largest off-diagonal weight which may be decimated
        \begin{equation}
            2 w_{ab}^2 = |M_{ab}|^2 + |M_{ba}|^2 - |M'_{ab}|^2 - |M'_{ba}|^2
            \label{eqn:off_weight}
        \end{equation}
        and conjugates \(M\) by the unitary \(R\) with a single non-trivial sub-block given by Eq.~\eqref{eqn:normal_R}. Repeating this step brings \(M\) eventually to a diagonal form~\cite{Goldstine1959}.
        
\begin{figure}
            \centering
            \includegraphics{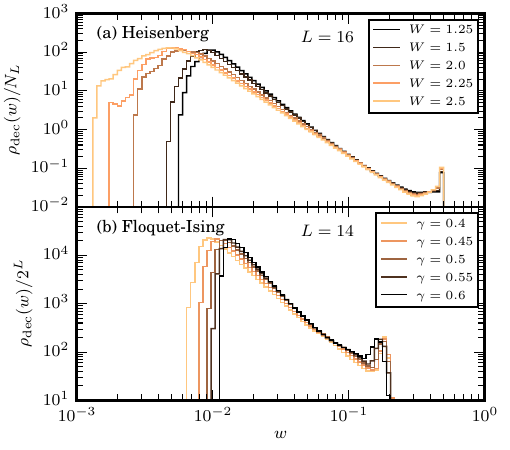}
            \caption{The number density of decimated elements for (\textbf{a}) the Heisenberg model~\eqref{eqn:Heis_model} (with Hilbert space dimension in the zero magnetization sector \(N_L = \binom{L}{L/2}\)) and (\textbf{b}) the Floquet Ising model~\eqref{eqn:FIsi_model} (with Hilbert space dimension \(2^L\)) show the same qualitative features as the Floquet circuit model~\eqref{eqn:model} (\autoref{fig:mat_dist}). We extract an estimate of \(\theta\) by using the maximum likleihood estimator~\eqref{eqn:MLE} with limits chosen to include the power-law segment where \(\rho_{\mathrm{dec}}\) scales with the Hilbert space dimension.}
            \label{fig:other_rho}
        \end{figure}

        As the algorithm progresses, we accumulate a histogram of the decimated weights \(w_{ij}\) (which is just \(|H_{ij}|\) in the Hermitian case). This is the histogram shown in \autoref{fig:mat_dist} for the Floquet circuit model. The distributions of decimated elements in the Heisenberg and Floquet-Ising models (both using the Hermitian algorithm) are shown in \autoref{fig:other_rho}.
        
        \begin{figure}
            \centering
            \includegraphics{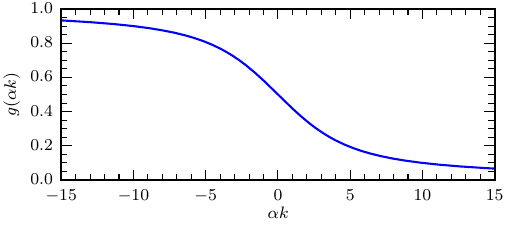}
            \caption{The left hand side in the equation~\eqref{eqn:MLE} defining the maximum likelihood estimator for \(\theta\).}
            \label{fig:MLE}
        \end{figure}
        
        To fit a power law to the distributions in \autoref{fig:mat_dist} and \autoref{fig:other_rho}, we use a maximum-likelihood estimator. Assuming that the numerically obtained decimated elements which lie between some cutoffs \(w_{\min}\) and \(w_{\max}\) are drawn from a power law distribution \(\rho_{\mathrm{dec}}(w) \propto w^{-1-\alpha}\) (so \(\theta = 1-\alpha\)), the most likely value for \(\alpha\) solves the equation~\cite{Clauset2009} (\autoref{fig:MLE})
        \begin{equation}
            k g(\alpha k) = \frac{1}{\alpha} - \frac{k}{e^{\alpha k}-1} = \frac{1}{N} \sum_{w \in [w_{\min}, w_{\max}]} \log \frac{w}{w_{\min}},
            \label{eqn:MLE}
        \end{equation}
        where \(k = \log(w_{\max}/w_{\min})\), and \(N\) is the number of data points in \([w_{\min}, w_{\max}]\). The estimator for \(\alpha\) (and hence \(\theta\)) is found by numerically solving this equation, with the right hand side measured from data.

\begin{figure}
            \centering
            \includegraphics[width=\linewidth]{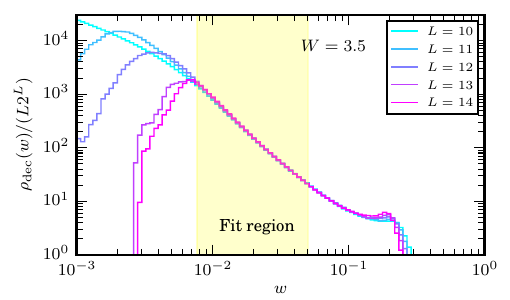}
            \caption{\label{fig:data_collapse}The histograms of decimated weights \(\rho_{\mathrm{dec}}\) for chains of different lengths \(L\) in the Floquet circuit model~\eqref{eqn:model} show data collapse when rescaling by \(L 2^L\). The region which is used to estimate \(\theta\) is shaded in yellow.}
        \end{figure}

        We choose the cutoffs \(w_{\min}\) and \(w_{\max}\) so that \(\rho_{\mathrm{dec}}(w)\) agrees with the proposed scaling form \(\rho_{\mathrm{dec}}(w) \propto N_L |H|^{-2+\theta}\) (where \(N_L\) is the Hilbert space dimension of a length \(L\) chain). In particular, we should obtain data collapse when rescaling \(\rho_{\mathrm{dec}}\) by \(N_L\). In fact, the collapse is improved when restoring one of the sub-exponential factors of \(L\) we have been neglecting in \(\rho_{\mathrm{dec}}\)~(\autoref{fig:data_collapse}). The cutoffs \(w_{\min}\) and \(w_{\max}\) are chosen to include this scaling regime.

        A more complete description of of \(\rho_{\mathrm{dec}}\) would likely include a flow of \(\theta\) with the flow time \(\Gamma\), as we comment in the main text. This should also be reflected by some dependence on the cutoff \(w_{\min}\). The prediction of stretched exponential decay assumes this flow is small enough to ignore. In our numerical fits, we use the smallest \(w_{\min}\) at which we have a numerical estimate of \(\rho_{\mathrm{dec}}\) and which lies within the \(\rho_{\mathrm{dec}} = O(N_L)\) regime. This corresponds to the longest times.
        
        The obtained estimates \(\theta\) are those shown in \autoref{fig:phase_diagram}(b). The system size used for the Floquet circuit and Floquet-Ising models is \(L=14\), while that for the Heisenberg model is \(L=16\). Errorbars are 70\% bootstrap confidence intervals obtained from 100 disorder realizations.

    \subsection{Fitting Autocorrelators}
        \label{subapp:fits}
        
        The procedure involved in fitting a stretched exponential to the numerically evaluated autocorrelators \(C(t)\) is, for the most part, straightforward. It uses a least squares optimization package. Less straightforward is the justification that stretched exponentials are the best functional form to fit. This section briefly summarizes the details of the fitting procedure, and then enumerates our reasons for preferring a stretched exponential fit as opposed to any other.
        
        As stated, we fit \(C(t)\) to a stretched exponential using standard least squares optimization. The only subtlety in the Floquet circuit model is that the fit is to logarithmically spaced points in time in the range \(t/T \in [6W, 2^L/(20 \pi)]\) so as not to bias the large \(t\) part of the curve. The lower cutoff grows with \(W\) to account for short time dynamics, and the upper cutoff is proportional to the Heisenberg time---except when the upper cutoff exceeds the maximum integration time \(10^4 T\). In that case, the cutoff is given by the maximum time. The error bars plotted in \autoref{fig:phase_diagram} and \autoref{fig:correlators}(b) are 70\% bootstrap confidence intervals obtained from between 200 and 1000 disorder realizations (depending on \(L\)).
        
        The fit to the data of Ref.~\cite{Lezama2019} on the Floquet-Ising model again uses logarithmically spaced data, with a lower cutoff chosen to avoid the initial short-time dynamics. The autocorrelator data for the Heisenberg model~\cite{Schiulaz2019} (\autoref{fig:Ct_Heis}) has a significant late time saturated value, due to the more significant finite-size effects of Hamiltonian models. In order to make use of the largest range of data, we must introduce a fourth fit parameter for the late time value, \(C(t) \sim A e^{-(t/\tau)^\beta}+B\).
        
        The more subtle point is determining that a stretched exponential is actually a good form to fit to the data. The first and most relevant reason to fit stretched exponential decay is the theoretical justification. As explained in the main text, such a functional form is natural when forming a hierarchy of resonances at different timescales. This theoretical candidate function shows excellent agreement with data (\autoref{fig:correlators} of the main text, and \autoref{fig:Ct_Heis}).
        
        On the other hand, other plausible empirical fits to the data of \autoref{fig:correlators}, including a logarithmic fit \(C(t) \sim A + B\log(t/T)\)~\cite{Sels2021} or a power law \(C(t) \sim A t^{-p}\) (not shown), fail to capture the behavior of \(C(t)\) over its entire range. Indeed, a curve in the \(C(t)\) data is visible even by eye when plotted on a log scale, except in the largest values of \(W\), where we do not have the dynamic range to distinguish a logarithm from a stretched exponential in any case.
        
        \begin{figure}
            \centering
            \includegraphics{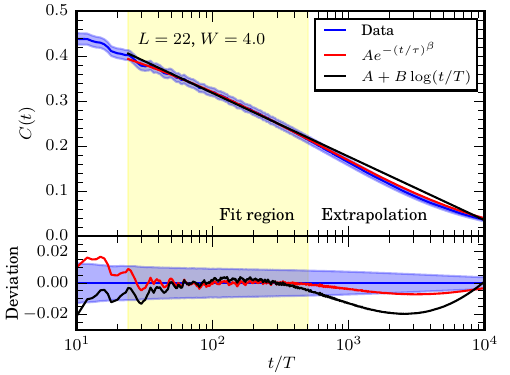}
            \caption{Fitting a stretched exponential to \(C(t)\) for the Floquet circuit model in a limited time window (yellow shaded) and extrapolating (red) almost reproduces the actual late time data (blue shaded region is one standard error of the mean). Even a decade in \(t/T\) beyond the fit region, the extrapolated curve is only slightly further than one standard deviation from the true value. A similar attempt in fitting a logarithm (black) produces a poor extrapolation, and a bend in the residuals even inside the fit region.}
            \label{fig:extrapolate}
        \end{figure}
        
        \autoref{fig:extrapolate} shows an example where a logarithmic fit seems plausible, but a stretched exponential performs notably better. In the highlighted range, the decay of \(C(t)\) seems logarithmic by eye. However, fitting a logarithm and inspecting the residuals reveals that even this limited range has a (small) bend in the data which a logarithm cannot account for. This manifests as a visible quadratic part in the residuals.
        
        Furthermore, we can also assess how well a logarithm and stretched exponential perform in accounting for data we did not fit. That is, we can attempt to extrapolate the fit, and check how well each candidate fit matches to late time data. Again, the stretched exponential outperforms a logarithm. Over more than a decade in \(t/T\), the stretched exponential only deviates from \(C(t)\) by slightly more than one standard deviation. The logarithm strays by closer to three.
        
        Extrapolating further clearly exacerbates this issue. The logarithmic fit eventually becomes negative, which \(C(t)\) never does.
        
        The point can reasonably be made that a stretched exponential fit uses more fit parameters than a logarithm---three rather than two. Nonetheless, this does not alleviate the fact that a logarithmic fit cannot account for all the features of \(C(t)\), and is disfavored by theory. Attempting to fit a power law in \autoref{fig:extrapolate} produces even worse results. Another candidate fit with three fit parameters could potentially outperform the stretched exponential, but we have not been able to find one, nor are any naturally provided by theory. Further, we note that introducing additional fit parameters usually makes extrapolation worse, so \autoref{fig:extrapolate} should not be expected to improve upon adding more parameters. As an example, fitting a quadratic in \(\log(t/T)\) to the data in \autoref{fig:extrapolate} produces even poorer extrapolations than a linear function in \(\log(t/T)\), as the data changes concavity.
        
        \begin{figure}
            \centering
            \includegraphics{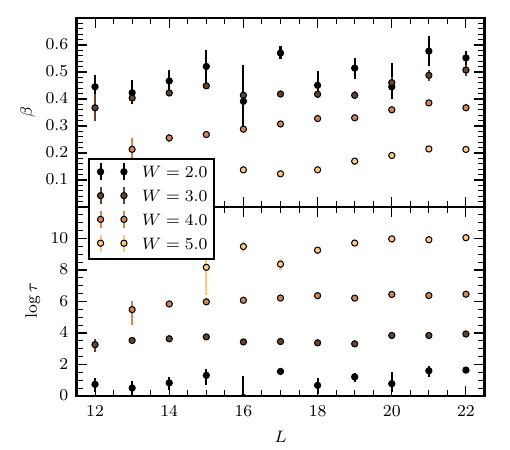}
            \caption{Fits of the stretch exponent \(\beta\) and time constant \(\tau\) to the data in \autoref{fig:correlators} show some drift in \(L\). For small disorder strengths \(W\), thermalization is too rapid to reliably fit a stretch exponent.}
            \label{fig:fits}
        \end{figure}
        
        Another important feature in assessing the fits is how the optimal fit parameters flow with \(L\). A large drift in the fit parameters would indicate a sensitive dependence on the late time data, where the largest \(L\) is needed to obtain convergence in \(C(t)\). The flow of the stretch exponent \(\beta\) and time constant \(\tau\) is shown in \autoref{fig:fits}. The time constant seems to not vary much as \(L\) is increased. The stretch exponent drifts slowly.

\begin{figure}
            \centering
            \includegraphics{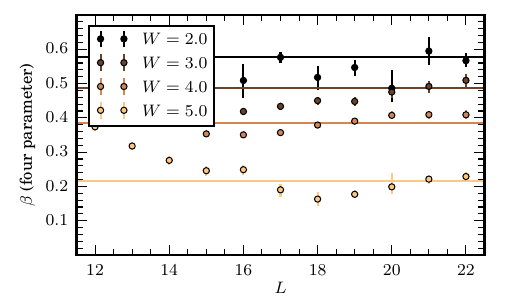}
            \caption{Fits of the stretch exponent \(\beta\) to the data in \autoref{fig:correlators} using a four parameter fit do not show the systematic drift of \autoref{fig:fits}. Solid lines mark the \(L=21\) values of \(\beta\) in the three parameter stretched exponential fit.}
            \label{fig:four_fits}
        \end{figure}

        Numerically, the flow of \(\beta\) with \(L\) is likely due to the non-zero late-time value of \(C(t)\) at small sizes affecting the fit. We cut off our fit range well before the Heisenberg time, but this feature may still have a systematic effect. Making a four parameter fit including a non-zero limit value, \(C(t) \sim A e^{-(t/\tau)^\beta} + B\), removes the systematic nature of the drift (\autoref{fig:four_fits}).

        Only a finite value of \(L\) is necessary to account for dynamics within a time \(O(\tau)\), so the flow of \(\beta\) should eventually saturate to a renormalized value. In the Jacobi picture, this may be due to a drift in \(\theta\) with the flow time. Our analysis does not account for such a drift.
        
        At small values of \(W\), the separation of scales between the initial drop of the correlator (\(O(W)\)) and the eventual equilibration time (\(O(\tau)\)) is too small to reliably fit a stretched exponential.

\section{Off-diagonal Matrix Elements}
    \label{app:offdiag}
    
    The distribution of off-diagonal matrix elements of a local operator reveals features of thermalization or localization. In the eigenbasis of a Hamiltonian satisfying the eigenstate thermalization hypothesis (ETH), off-diagonal matrix elements of local operators (other than the Hamiltonian itself) are normally distributed with an \(\omega\)-dependant variance~\cite{Srednicki1994,DAlessio2016}. Meanwhile, in a quasilocal basis, the distribution has a power law component for small matrix elements, with an exponent related to the localization length~\cite{Crowley2020}.
    
\begin{figure}
        \centering
        \includegraphics{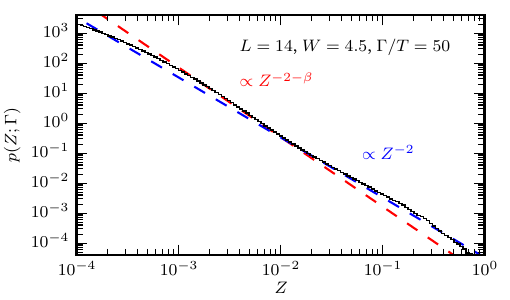}
        \caption{The distribution of off-diagonal matrix elements in the Jacobi basis \(Z_{ab}(\Gamma) = |\bra{a(\Gamma)}\sigma^z_0\ket{b(\Gamma)}|\) shows two power law segments. At large \(Z\), the distribution has a \(Z^{-2}\) tail (blue dashed). For small \(Z\) there is a \(Z^{-2-\beta}\) regime (red dashed). This double power law is predicted by Eq.~\eqref{eqn:rhoZ_plaw}. The small \(Z\) cut off is related to the Jacobi flow time. The kink separating these regimes moves towards small \(Z\) as the Jacobi flow time \(\Gamma\) is increased. Data is from the Floquet circuit model, and \(\beta\) is fit from data in \autoref{fig:correlators}.}
        \label{fig:offdiag}
    \end{figure}

    What of the matrix elements \(Z_{ab}(\Gamma) = |\bra{a(\Gamma)}\sigma^z_0\ket{b(\Gamma)}|\) in the dressed Jacobi basis \(\{\ket{a(\Gamma)}\}\)? A simplified calculation of \(p(Z;\Gamma)\), the distribution of off-diagonal elements \(Z_{ab}(\Gamma)\), predicts this distribution also develops a small matrix element power law component, with the exponent being \(-2+\theta = -2-\beta\). We also predict an intermediate regime where a different power law emerges with an exponent \(-2\), in plausible agreement with our numerics~(\autoref{fig:offdiag}), and previous results in the literature~\cite{Garratt2021,Garratt2022}.
    
    Each of the power law segments can be related to a different feature of the Jacobi algorithm: the \(-2+\theta\) segment arises from the distribution of small decimated elements, while the \(-2\) segment corresponds to the presence of resonances.
    
    The key simplifying assumptions that make the simplified calculation tractable are that resonances are sparse, and that the distribution of decimated elements is a power law.
    
    In the initial basis \(\{\ket{a(\Gamma_0)}\}\) the operator \(\sigma^z_0\) is diagonal, with elements \(\pm 1\). Thus, if the Jacobi procedure makes a rotation between the states \(\ket{a}\) and \(\ket{b}\) with nontrivial block
    \begin{equation}
        R_0 = 
        \begin{pmatrix}
            \cos(\eta_0/2) & -e^{-i \phi_0} \sin(\eta_0/2) \\
            e^{i \phi_0} \sin(\eta_0/2) & \cos(\eta_0/2)
        \end{pmatrix},
        \label{eqn:rot}
    \end{equation}
    (with \(\eta_0 \in [0,\pi]\)) in the first time step, then the off-diagonal matrix elements are
    \begin{equation}
        Z_{ij}(\Gamma_1) = |\bra{i} R_0^\dagger \sigma^z_0 R_0 \ket{j}| = (\delta_{ai} \delta_{bj} + \delta_{aj} \delta_{bi}) \sin\eta_0,
    \end{equation}
    provided \(\qexp{\sigma^z_0}{a} \neq \qexp{\sigma^z_0}{b}\).
    
    We make the simplifying assumption that resonances are sparse, so that elements of this form are all that appear in the rotated \(\sigma^z_0\). That is, we neglect the effect of future rotations on the elements \(\sin\eta_k\) generated at time \(\Gamma_k\), and treat the distribution \(p(Z;\Gamma)\) as measuring the density of the values \(\sin\eta_k\) for \(\Gamma_k \leq \Gamma\). This approximation provides the expression
    \begin{equation}
        p(Z; \Gamma) = N\int \d H\d\eta\, \delta(Z - \sin\eta) p(\eta \mid H) \rho_{\mathrm{dec}}(H; \Gamma),
        \label{eqn:rhoZ_int}
    \end{equation}
    where \(N\) is a normalization constant, \(p(\eta \mid H)\) is the distribution of rotation angles given that the decimated element is \(H\), and \(\rho_{\mathrm{dec}}(H; \Gamma)\) is the number density of Hamiltonian elements decimated before flow time \(\Gamma\).
    
    We assume the density of decimated elements \(\rho_{\mathrm{dec}}(H; \Gamma)\) is given by a power law (Figs.~\ref{fig:mat_dist},~\ref{fig:other_rho}), as in the main text:
    \begin{equation}
        \rho_{\mathrm{dec}}(H; \Gamma) = 2^L C |H|^{-2 + \theta} 1_{[H_{\min}(\Gamma), J]}(H).
        \label{eqn:rhoH_cut}
    \end{equation}
    Above \(1_{A}\) is the indicator function for the set \(A\). The lower cutoff \(H_{\min}(\Gamma)\) decreases with flow time as more elements are decimated. The upper cutoff \(J\) is set by the initial energy scale of the off-diagonal part of the Hamiltonian in the product state basis.
    
    Given a value for the decimated element \(H\), the rotation angle is determined (in the Hamiltonian case, for brevity) by the difference in diagonal elements \(\omega\),
    \begin{equation}
        t=|\tan\eta_k| = \frac{|H|}{|\omega|/2}.
        \label{eqn:taneta}
    \end{equation}
    If the distribution \(p(\omega)\) has a finite value at \(\omega=0\), then the distribution of the quotient \(p(x=1/|\omega|)\) acquires a heavy \(c x^{-2}\) tail for large \(x\), where \(c = 2p(\omega=0)\). The exact form of this distribution can be calculated through a change of variables given some \(p(\omega)\), but we will again make a simplifying (though inessential) assumption that the distribution is a pure power law, with a lower cutoff determined by normalization.
    \begin{equation}
        p(x = 1/|\omega|) = c x^{-2} 1_{[c,\infty)}(x)
    \end{equation}
    From this expression, finding the probability density of \(t= |\tan\eta|\) just involves scaling by a constant factor (assuming \(\omega\) and \(H\) are uncorrelated).
    \begin{equation}
        p(t \mid H) = \frac{c}{2|H|} \left( \frac{t}{2|H|} \right)^{-2} 1_{[c,\infty)}\left( \frac{t}{2|H|} \right)
        \label{eqn:prob_taneta}
    \end{equation}
    The delta function in Eq.~\eqref{eqn:rhoZ_int} enforces \(Z = \sin\eta =: s\). Re-writing \(p(t\mid H)\) in terms of \(s\):
    \begin{equation}
        p(s \mid H) = \frac{ 2c |H| s^{-2}}{(1-s^2)^{1/2}} 1_{[c,\infty)}\left( \frac{s}{2|H|(1-s^2)^{1/2}} \right).
        \label{eqn:prob_sineta}
    \end{equation}
    
    Substituting Eqs.~\eqref{eqn:rhoH_cut} and \eqref{eqn:prob_taneta} into Eq.~\eqref{eqn:rhoZ_int} and integrating the delta function gives, defining \(z = Z/\sqrt{1-Z^2}\),
    \begin{align}
        p(Z; \Gamma) &= N\int \d H\d s\, \delta(Z - s) p(s\mid H) \rho_{\mathrm{dec}}(H; \Gamma), \nonumber \\
        &= \frac{2^{L+1} N c C z^{-2}}{(1-Z^2)^{3/2}} \int_{H_{\min}(\Gamma)}^J\d H\,  H^{-1+\theta} 1_{[c,\infty)}(\tfrac{z}{2H}).
    \end{align}
    There is also an upper cutoff \(Z \leq 1\), as \(Z = \sin\eta\). We split the remaining \(H\) integral into two cases,
    \begin{multline}
        \int_{H_{\min}(\Gamma)}^J\d H\,  H^{-1+\theta} 1_{[c,\infty)}(\tfrac{z}{2H}) \\
        = \left\{
        \begin{array}{l l}
            \left(J^\theta - H_{\min}(\Gamma)^\theta\right)/\theta & z \geq 2 c J, \\
            \left((z/2c)^\theta - H_{\min}(\Gamma)^\theta\right)/\theta & z < 2 c J.
        \end{array}
        \right.
        \label{eqn:rhoZ_plaw}
    \end{multline}
    When \(z < 2 c J\), the distribution of matrix elements is a power law \(\propto z^{-2 + \theta}\). For small \(Z \ll 1\), this is essentially a \(Z^{-2+\theta}\) distribution. This segment originates from the distribution of decimated elements, and for \(\theta >0 \) is characteristic of localized systems~\cite{Crowley2020}. 
    
    When \(z\) is larger than \(2cJ = O(J/W)\), but still much less than 1, there is a \(z^{-2} \approx Z^{-2}\) power law tail, due to the heavy tail in the \(\tan\eta\) distribution. The origin of that tail was small energy differences \(\omega\)---that is, resonances. As this power law only emerges for \(2cJ \ll Z \ll 1\), there must be a sepration of scales \(cJ \ll 1\). In the bare Hamiltonian \(c= O(1/W)\), and only decreases throughout the course of the Jacobi algorithm, so \(2cJ = O(J/W)\). The separation \(cJ \ll 1\) is expected when \(W/J\) is large.
    
    This two-segment power law is reflected in numerics, as shown in the plot of \(p(Z;\Gamma)\) in \autoref{fig:offdiag}. There are two power law components with roughly the right exponents, separated by a kink, as predicted by Eq.~\eqref{eqn:rhoZ_plaw}.
    
    The kink separating the two power law regimes moves towards smaller \(Z\) as the flow time is increased (data not shown). This feature is not accounted for with our simplified calcualtion, and presumably requires a more detailed treatment to understand.
    
    The two-segment power law is visible elsewhere in the literature. For instance, in Refs.~\cite{Garratt2021,Garratt2022} both the large matrix element \(\theta\) independent part and the turnover into a \(\theta\) dependent part are visible, though those references focused on the \(Z^{-2}\) tail associated to resonances. Ref.~\cite{Garratt2022} also argued that \(Z = |\Omega/\omega|\), where \(\Omega\) is uncorrelated with \(\omega\). In the current analysis, we see that this is a consequence of Eq.~\eqref{eqn:taneta} in the \(\eta \ll 1\) limit, where \(Z = s \approx t = |2H/\omega|\).

\section{Successive Resonances and Dynamics}
    \label{app:multires}
    
    This appendix describes details of the calculation predicting stretched exponential decay of autocorrelators in the resonance model, Eq.~\eqref{eqn:Ct_prediction}. The distribution of matrix elements which cause a resonance is characterized in Appendix~\ref{subapp:res}---in particular, the lower cutoff to the power law regime is estimated. The stretched exponential form of the decay, and the associated timescale, is derived in Appendix~\ref{subapp:decay_form}.
    
    \subsection{The Distribution of Resonances}
        \label{subapp:res}
    
        The distribution of resonance frequencies, \(\rho_{\mathrm{res}}(|H|)\), is a vital ingredient in determining the form of autocorrelator decay. In this section, we fill in details regarding the characterization of \(\rho_{\mathrm{res}}(|H|)\) not included in the main text. The cutoffs within which
        \begin{equation}
            \rho_{\mathrm{res}}(|H|) = 2^L C |H|^{-2+\theta}
            \label{eqn:rhodec_pow}
        \end{equation}
        (assuming a Hilbert space dimension of \(2^L\) for a length \(L\) chain) will play an important role in setting the timescale for the decay of \(C_Z(t)\).
        
        As in the main text, we suppose the number density of decimated elements is given by the power law~\eqref{eqn:rhodec_pow}, and thus that the number density of elements responsible for a resonance is also a power law,
        \begin{equation}
            \rho_{\mathrm{res}}(|H|) = 2^L c C |H|^{-1 + \theta}.
            \label{eqn:rho_res2}
        \end{equation}
        Here, \(c = 2p(\omega=0)\) is the probability density for \(|\omega|\) at zero frequency, as in Appendix~\ref{app:offdiag}, and \(C=O(W^0)\) is the coefficient appearing in the full distribution of decimated elements.
        
        As matrix elements cannot be arbitrarily large, there is an upper cutoff to this density set roughly by the local bandwidth \(J\). Further, beyond some maximum flow time \(\Gamma_{\max} = 2\pi/\omega_c\) for the Jacobi algorithm, resonances can no longer be considered to be sparse. This manifests in \(\rho_{\mathrm{res}}\) as a growth with \(L\) which is faster than \(2^L\). Worst case estimates for the convergence of the Jacobi algorithm predict that \(\rho_{\mathrm{res}}\) cannot grow faster than \(4^L\)~\cite{Golub2000}. The estimate of \(4^L\) arises from assuming that all matrix elements \(H_{ab}\) are of the same scale. The \(2^L\) scaling assumes that an \(O(1)\) number of matrix elements dominate the norm of a row---which is the regime of sparse resonances.
        
        Indeed, physically, the lower cutoff \(\omega_c\) reflects the breakdown of the assumption that dynamics is due to sparse resonances. Eventually, the Jacobi procedure must begin to significantly affect the same states \(\ket{a}\) many times. At this point, the distinction between what is called a resonance and what is not begins to break down, and eventually ETH statistics begin to emerge. As such, we suppose that \(\omega_c\) is set by the condition that \(a = O(1)\) resonances have been encountered per state:
        \begin{equation}
            \int_{\omega_c}^J \d H\, \rho_{\mathrm{res}}(H) = a 2^L
            \implies
            \left(\frac{\omega_c}{J}\right)^\theta = 1 -\theta\frac{a}{cC J^\theta}.
            \label{eqn:wc1}
        \end{equation}
        The cutoff \(\omega_c\) is strictly positive when \(\theta <0\). If \(\theta>0\) then it is possible \(\omega_c =0\). 

        Note that the factor
        \begin{equation}
            c C J^\theta = O(J/W)
            \label{eqn:cCJ}
        \end{equation}
        by dimensional analysis of \(C\) and \(c = O(W^{-1})\).

    \subsection{Form of the Decay}
        \label{subapp:decay_form}
    
        In order to evaluate the autocorrelator \(C_Z(t)\) from Eq.~\eqref{eqn:Ct_char_fn}, we need to find \(p_{\mathrm{res}}(\omega)\), or rather, its Fourier transform. In fact, it is easier to compute the logarithm of the Fourier transform directly, rather than \(p_{\mathrm{res}}(\omega)\). In this section, we carry out the computation of \(C_Z(t)\), finding it decays as a stretched exponential~\eqref{eqn:Kt}, and further that the decay time is exponentially small in disorder~\eqref{eqn:theta_decay_W}.
        
        Recall that \(p_{\mathrm{res}}(\omega)\) is the distribution function for the sum
        \begin{equation}
            \omega = \sum_{k} \mu_k H_k
            \label{eqn:omega_sum}
        \end{equation}
        with \(\mu_k = \pm 1\), and each \(H_k\) drawn independently from \(\tilde{\rho}_{\mathrm{res}} = 2^{-L} \rho_{\mathrm{res}}\), so that there are on average \(a\) terms in the sum (Appendix~\ref{subapp:res}). As alluded above, it is easier to directly compute the cumulant generating function (CGF) \(K(t)\) defined by
        \begin{equation}
            p_{\mathrm{res}}(\omega) = \int \d t \, \exp(-i \omega t + K(t)).
        \end{equation}
        Comparing to Eq.~\eqref{eqn:Ct_char_fn}, 
        \begin{equation}
            K(t) = \log C_Z(t) - \log(2^L[Z^2]_{p_\mathrm{res}}).
        \end{equation}
        
        The sum~\eqref{eqn:omega_sum} can be reformulated as an integral over \(H\). Consider windows \([H, H + \d H)\) and associate to each a random variable \(h\) which takes the values \(\{H,-H,0\}\) with respective probabilities
        \begin{equation}
            \{\tfrac{1}{2}\tilde{\rho}_{\mathrm{res}}(H)\d H,\tfrac{1}{2}\tilde{\rho}_{\mathrm{res}}(H)\d H,1-\tilde{\rho}_{\mathrm{res}}(H)\d H\}.
        \end{equation}
        Then \(\omega\) is the sum of the \emph{independent} random variables \(h\). The CGF of this (discrete) random variable is
        \begin{multline}
            K_h(t) = \log\left\{ 1-\tilde{\rho}_{\mathrm{res}}(H) \d H\right. \\
            \left. + \tfrac{1}{2}\tilde{\rho}_{\mathrm{res}}(H)\d H (e^{-i H t} + e^{+i Ht} )\right\}
        \end{multline}
        which simplifies to
        \begin{equation}
            K_h(t) = (-1 + \cos(Ht)) \tilde{\rho}_{\mathrm{res}}(H) \d H
        \end{equation}
        with \(\tilde{\rho}_{\mathrm{res}}(H) \d H \ll 1\).
    
        Using the fact that the CGF of a sum of independent random variables is the sum of the CGFs of the summands, we have
        \begin{align}
            K(t) &= \int K_h(t) \nonumber \\
            &= -a + \int \d H\, \tilde{\rho}_{\mathrm{res}}(H) \cos(H t).
        \end{align}
        The integral is (the real part of) the Fourier transform of \(\tilde{\rho}_{\mathrm{res}}(H)\). The power law form~\eqref{eqn:rho_res2} gives
        \begin{equation}
            K(t) \sim -(t/\tau)^{-\theta},\quad J^{-1} \ll t \ll \omega_c^{-1}
            \label{eqn:Kt}
        \end{equation}
        for some scale \(\tau\).
        
        Recalling that \(C_Z(t) = 2^L [Z^2]_{p_{\mathrm{res}}} e^{K(t)}\), Eq.~\eqref{eqn:Kt} is the predicted stretched exponential form of the autocorrelator.
        
        To calculate the time scale \(\tau\), we simplify with hard cutoffs for \(\tilde{\rho}_{\mathrm{res}}(|H|)\). Then \(K(t)\) can be computed exactly in terms of special functions. For \(\theta < 0\),
        \begin{multline}
            \int_x^\infty \d H\, H^{-1+\theta} \cos(H t)
            \\= t^{-\theta} \Gamma(\theta) \cos(\tfrac{\pi\theta}{2})
            + \tensor[_1]{F}{_2}(\tfrac{\theta}{2};\tfrac{1}{2},1+\tfrac{\theta}{2};-(\tfrac{xt}{2})^2),
        \end{multline}
        where \(\tensor[_1]{F}{_2}(a_1;b_1,b_2;z)\) is a hypergeometric function, which we subsequently abbreviate as
        \begin{equation}
            f_\theta(xt) = \tensor[_1]{F}{_2}(\tfrac{\theta}{2};\tfrac{1}{2},1+\tfrac{\theta}{2};-(\tfrac{xt}{2})^2).
        \end{equation}
        
        Including the normalization of \(\tilde{\rho}_{\mathrm{res}}\) (determined by the cutoffs and the number of resonances per state \(a\)), the CGF is
        \begin{align}
            K(t)/a &= -1 + \frac{\theta}{J^\theta - \omega_c^\theta}\int_{\omega_c}^J \d H\, H^{-1+\theta} \cos(H t) \\
            &= -1 + \frac{f_\theta(\omega_c t) - (\tfrac{\omega_c}{J})^{-\theta} f_\theta(Jt) }{1- (\tfrac{\omega_c}{J})^{-\theta}}.
        \end{align}
        
        Of more interest than the exact expression is the asymptotic expansion in the regime \(\omega_c t \ll 1 \ll Jt\), given by
        \begin{equation}
            K(t) = \frac{a}{1-(\tfrac{\omega_c}{J})^{-\theta}} - (t/\tau)^{-\theta} + O((Jt)^{-1}, (\omega_c t)^2),
            \label{eqn:hard_asymp}
        \end{equation}
        where
        \begin{equation}
            J \tau = \left(\frac{(\tfrac{J}{\omega_c})^{-\theta}-1}{a \theta \Gamma(\theta) \cos(\tfrac{\pi\theta}{2})} \right)^{-1/\theta},
            \label{eqn:tau_full}
        \end{equation}
        which is positive for \(\omega_c < J\) and \(\theta > -2\). (When \(\theta < -2\), the quadratic term in Eq.~\eqref{eqn:hard_asymp} becomes the dominant one.)

        Substituting  Eq.~\eqref{eqn:wc1} gives an expression for \(\log(J\tau)\) which depends on \(\theta\):
        \begin{multline}
            \log(J \tau) = -\frac{1}{\theta}\log\left( \frac{-1}{cC J^\theta \Gamma(\theta) \cos(\tfrac{\pi\theta}{2})} \right) \\
            = O\left(-\theta^{-1}\log(-\theta W/J)  \right).
            \label{eqn:theta_decay}
        \end{multline}
        If \(-\theta^{-1}\) has a linearizable dependence on \(W/J\) in some regime of interest~\cite{Crowley2020}, this gives
        \begin{equation}
            \log(J \tau) = O(W/J).
            \label{eqn:theta_decay_W}
        \end{equation}
        The time constant for stretched exponential decay grows exponentially in disorder, as we wanted to explain.
        
        Essentially identical calculations may be carried out in the localized phase, \(\theta > 0\). In that case, one finds
        \begin{equation}
            C_Z(t) \sim A e^{(\tau/t)^\theta} = A\left(1 + \left(\frac{\tau}{t}\right)^\theta + O((\tau/t)^{2\theta}) \right).
            \label{eqn:Ct_postheta}
        \end{equation}
        At late times \(t \gg \tau\), the autocorrelator approaches a constant as a power law. This is consistent with existing predictions on the form of \(C(t)\) in MBL~\cite{Serbyn2014}. The notable difference in Eq.~\eqref{eqn:Ct_postheta} is that the full expression (before expanding the exponential) has a time scale \(\tau\), unlike purely power law decay.

\end{document}